\documentclass[preprintnumbers,prd,floatfix,nofootinbib]{revtex4}

\pdfoutput=1 

\usepackage[tbtags]{amsmath}  
\usepackage{amssymb}          
\usepackage{bm}               
\usepackage{graphicx}         
\usepackage[export]{adjustbox}
\usepackage{hhline,multirow}  
\usepackage{dcolumn}          
\usepackage{slashed}
\usepackage{datetime}
\usepackage{color}
\usepackage[dvipsnames]{xcolor}

\usepackage{amscd}            
\usepackage{epsfig}
\usepackage{dcolumn}          
\usepackage[dvipsnames]{xcolor}
\usepackage[normalem]{ulem}
\usepackage{mathtools}

\newcommand{\vp}{v^{\,\prime}}
\newcommand{\vpsq}{v^{\,\prime\,2}}

\newcommand{\nbar}{{\overline n}}
\newcommand{\nslash}{n\hspace*{-0.22cm}\slash\hspace*{0.022cm}}
\newcommand{\nbslash}{\nbar\hspace*{-0.22cm}\slash\hspace*{0.022cm}}
\newcommand{\kslash}{k\hspace*{-0.22cm}\slash\hspace*{0.022cm}}
\newcommand{\kpslash}{k\hspace*{-0.22cm}\slash\hspace*{0.022cm}'}

\newcommand{\Tr}{{\rm Tr}}

\allowdisplaybreaks[2]

\begin{document}


\preprint{JLAB-THY-17-2596}

\title{Gauge invariance and kaon production in deep inelastic scattering at low scales} 

\author{Juan V. Guerrero}\email{juanvg@jlab.org}
\author{Alberto Accardi}\email{accardi@jlab.org}
\affiliation{
Hampton University, Hampton, VA 23668, USA}
\affiliation{
Jefferson Lab, Newport News, VA 23606, USA 
}


\begin{abstract}
This paper focuses on hadron mass effects in calculations of semi-inclusive kaon production in lepton-Deuteron deeply inelastic scattering at HERMES and COMPASS kinematics. In the collinear factorization framework, the corresponding cross section is shown to factorize, at leading order and leading twist, into products of parton distributions and fragmentation functions evaluated in terms of kaon- and nucleon-mass-dependent scaling variables, and to respect gauge invariance. It is found that hadron mass corrections for integrated kaon multiplicities sizeably reduce the apparent large discrepancy between measurements of $K^+ + K^-$ multiplicities performed by the two collaborations, and fully reconcile their $K^+/K^-$ ratios.
\end{abstract}



\maketitle

\section{Introduction}

During the last decade, significant advances have been made in the understanding of the partonic structure of the nucleon \cite{Jimenez-Delgado:2013sma,Gao:2017yyd}.
Currently, the valence quark and the gluon sectors are well understood, for which sets of Parton Distribution Functions (PDFs) extracted from a global data set are available with small uncertainties, except at large values of the parton momenta relative to the nucleon.
This is not the case, however, in the sea quark sector, for which the PDFs are less well known, and in particular in the strange sector. The strange quark PDF has been extracted phenomenologically in global QCD fits by several groups \cite{Ball:2014uwa,Harland-Lang:2014zoa,Dulat:2015mca,Alekhin:2017kpj}, largely relying on data on di-muon production in neutrino-nucleus scattering \cite{Goncharov:2001qe,Samoylov:2013xoa}, as well as from data on weak boson production in proton-proton collisions by the ATLAS and CMS experiments at the LHC \cite{Aad:2012sb,Chatrchyan:2013mza,Chatrchyan:2013uja,Aad:2014xca}. It has also been extracted experimentally from Semi-Inclusive Deeply Inelastic Scattering (SIDIS) data in kaon production by the HERMES collaboration \cite{Airapetian:2008qf,Airapetian:2013zaw}, and, with decreased sensitivity, from pion production data \cite{Yang:2015avi}. All of these show large differences in the size and shape of the obtained $s$-quark momentum distribution. Furthermore, tensions between nuclear target data and proton measurements at the LHC have been highlighted and discussed in Refs.~\cite{Alekhin:2014sya,Accardi:2016muk,Alekhin:2017olj}.

The strange quark PDFs can be separated from other flavors, {\it e.g}, by tagging kaons in SIDIS reactions and analyzing their multiplicity integrated over the kaon's momentum. These have been measured on Deuteron targets by the HERMES \cite{Airapetian:2013zaw,Airapetian:2012ki} and COMPASS \cite{Seder:2015sdw,Adolph:2016bwc} collaborations, that however show large discrepancies in their results. These measurements are sensitive to relatively low values of photon virtualities $Q^2$, where the mass of the target and observed hadron, generically denoted by $m$, induce ``Hadron Mass Corrections'' (HMCs) of order ${\mathcal{O}(m^2/Q^2)}$ that can be larger than the experimental uncertainties \cite{Accardi:2009md,Guerrero:2015wha}.
Crucially, with a kaon mass $m_K\approx 0.5$ GeV and scales $Q \approx 1-4$ GeV, hadron mass corrections may be non negligible even at relatively high energy experiments such as HERMES and COMPASS.

In this paper, we quantify HMC effects in HERMES and COMPASS data with calculations based on the formalism developed in Refs.~\cite{Accardi:2009md,Guerrero:2015wha}; this is recalled in Section~\ref{sec:formalism}, where we pay special attention to the conceptual underpinnings of the formalism and to explicitly discuss the gauge invariance of the mass-corrected SIDIS cross section, that has been criticized in Ref.~\cite{Christova:2016hgd}. In Section~\ref{sec:multiplicities} and \ref{sec:ratios}, we argue that HMCs are indeed not negligible, and may largely -- although not solely -- explain the observed differences between the measurements performed by the two collaborations (preliminary results were presented in Ref.~\cite{Guerrero:2017dcc}). This is especially true for the $K^+/K^-$ multiplicity ratios examined in Section~\ref{sec:ratios}, in which effects neglected in this analysis, such as Next-to-Leading Order (NLO) and Higher-Twist (HT) corrections, can be expected to largely cancel, as briefly discussed in Section~\ref{sec:otherQ2}. In Section~\ref{sec:conclusions} we summarize our findings and discuss prospects for future theoretical and phenomenological work, and in Appendix~\ref{app:baryon_number} we discuss in some detail our treatment of baryon number conservation.

\section{Leading order multiplicities at finite $Q^2$}
\label{sec:formalism}

\begin{figure}[bt]
	\centering
	\parbox[c]{5.5cm}{\includegraphics[width=\linewidth]{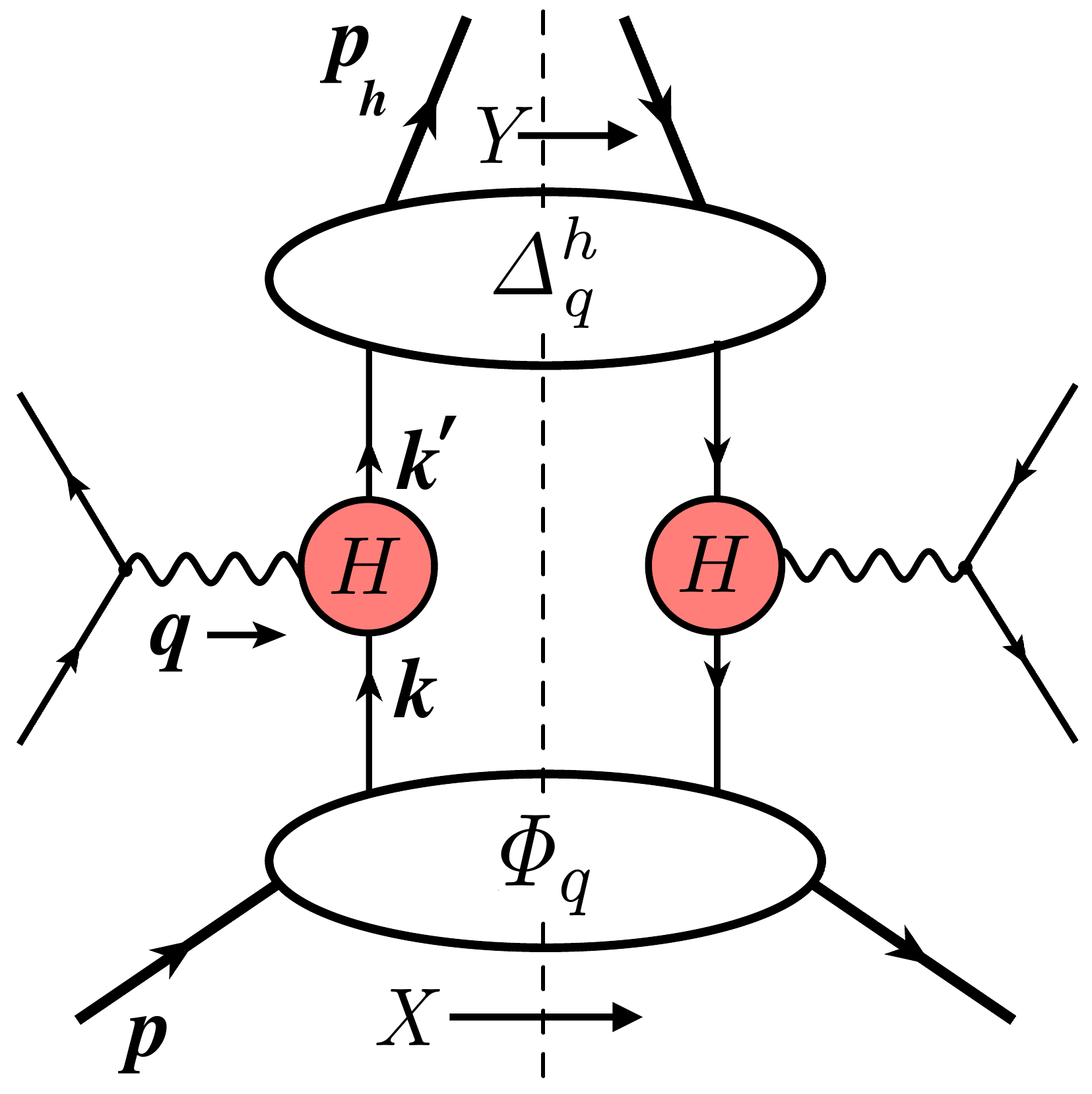}}
	\hspace{1cm}
	\parbox[c]{5.5cm}{\includegraphics[width=\linewidth]{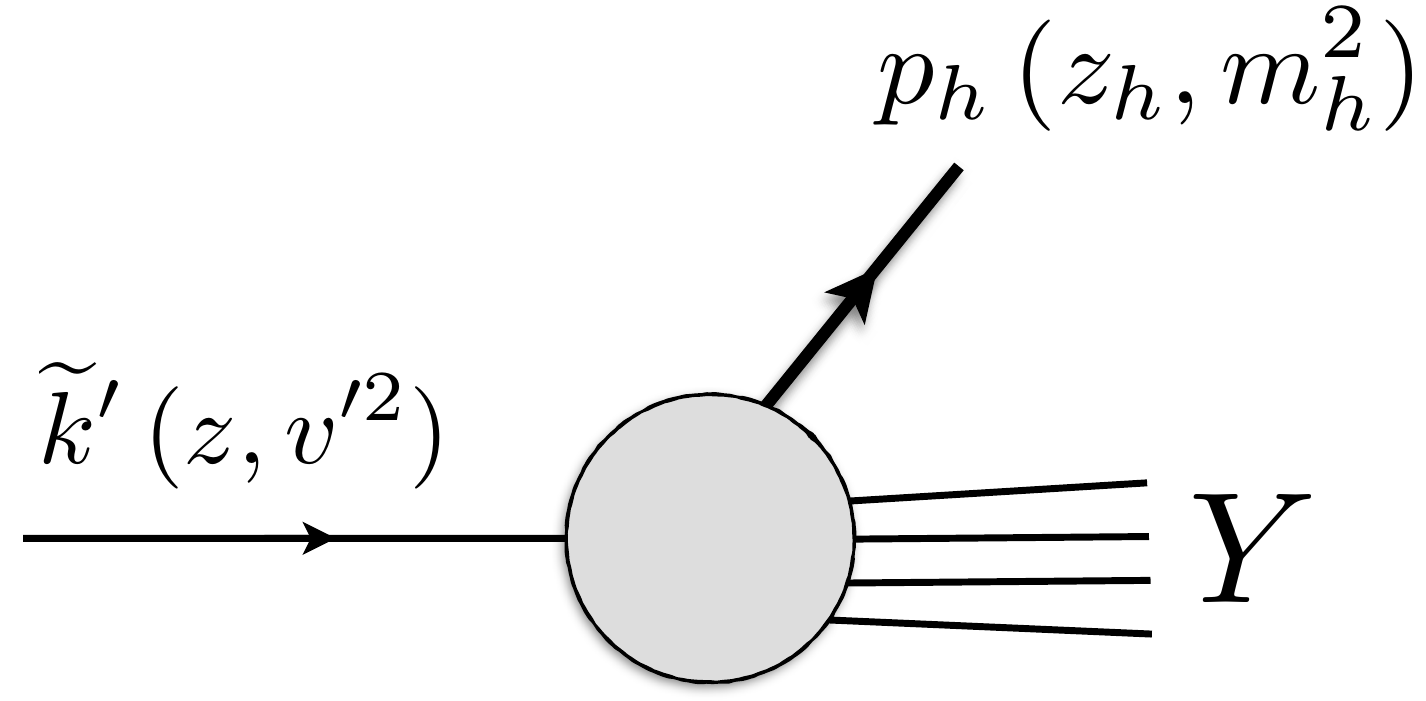}\\[.5cm]}     
	\caption{{\it Left:} SIDIS handbag diagram and kinematics, with $q$ the momentum of the photon, $p$ of the target nucleon, $p_h$ of the observed hadron, $k$ and $k'$ of the partons participating in the hard scattering $H$.
        {\it Right:} factorized kinematics at the fragmentation vertex, with $\widetilde{k'}$ the collinear, approximated fragmenting quark momentum.}
	\label{fig:SIDIS_handbag}
\end{figure}

The $z$-integrated hadron multiplicities measured by the HERMES collaboration \cite{Airapetian:2013zaw,Airapetian:2012ki} are defined as a ratio of the semi-inclusive to inclusive cross sections,
\begin{equation}
M^{h}(x_B^{\textrm exp}) = \frac{\int_{\textrm exp} dx_B dQ^2 \int_{0.2}^{0.8(0.85)}\, dz_h\, \frac{d\sigma^h}{dx_B dQ^2 dz_h}}{\int_{\textrm exp} dx_B dQ^2  \frac{d\sigma^{\textrm{DIS}}}{dx_B dQ^2}} \ ,
\label{eq:def_multiplicities}
\end{equation}
where $x_B= \frac{Q^2}{2 p \cdot q}$ is the Bjorken scaling variable, $Q^2 = -q^2$ the virtuality of the exchanged photon, $z_h= \frac{p \cdot p_h}{p \cdot q}$ is the fragmentation invariant, and the remaining kinematic variables are defined in Fig.~\ref{fig:SIDIS_handbag} left. The COMPASS collaboration has  measured integrated multiplicities as averages over $y$ of the differential ones, $\int dz_h \langle M^h(x_b,y,z_h)\rangle_y$ \cite{Seder:2015sdw,Adolph:2016bwc}; however, since $y<0.7$ within the COMPASS kinematic cuts, the two definitions are approximately equivalent, and in this paper we will use Eq.~\eqref{eq:def_multiplicities} for both experiments.

\begin{figure}[bt]
	\centering
	\includegraphics[width=\linewidth]{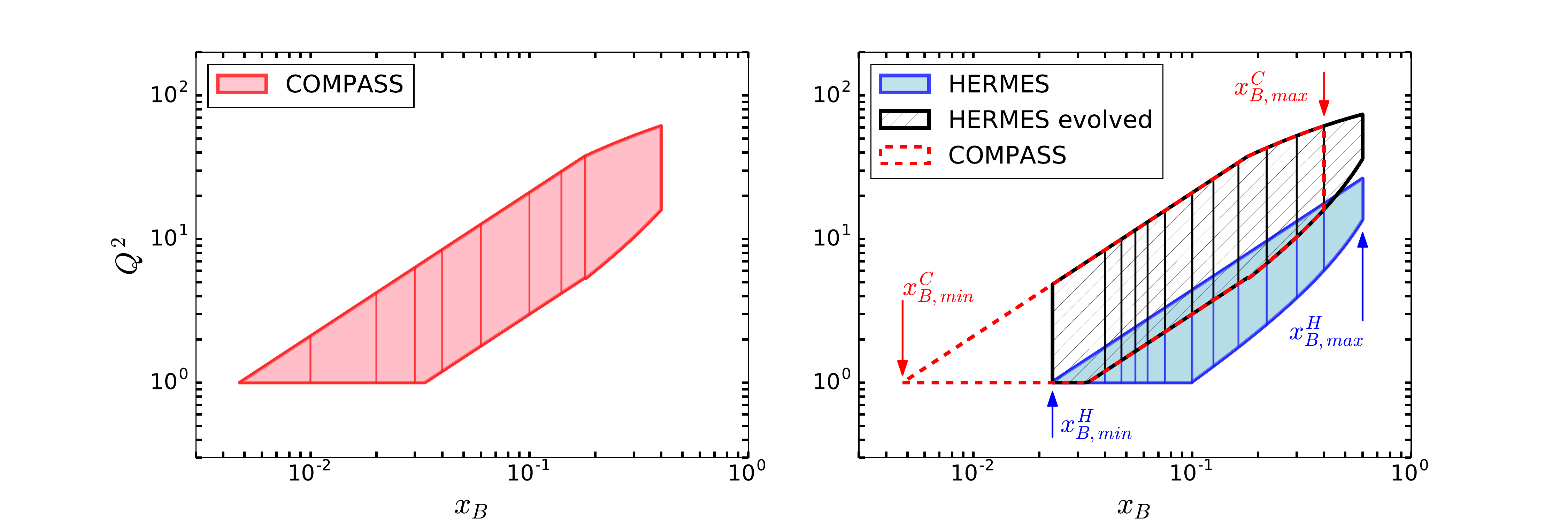}    
	\caption{{\it Left:} COMPASS experimental kinematic acceptance. {\it Right:} HERMES experimental kinematic acceptance (blue) and HERMES experimental binning evolved at COMPASS spectrometer (black hatched).}
	\label{fig:H_C_PS}
\end{figure}

The integration over the initial state DIS invariants is performed over the experimental kinematic acceptance of each measurement~\cite{Seder:2015sdw,Aschenauer:2015rna}, denoted in short by ``exp''. In more detail, the integral over $dx_B$ is performed over the bin of nominal value $x_B^{\textrm exp}$, and the integration over $dQ^2$ is performed within $x_B$-dependent limits defined by the kinematic cuts of each experiment. 
These cuts, as well as plots of the $(x_B,Q^2)$ phase space with the experimental $x_B$ bins are shown in Fig.~\ref{fig:H_C_PS}. The integration limits on $z_h$ are explicitly indicated in Eq.~(\ref{eq:def_multiplicities}), with the COMPASS value in parentheses when different from that used at HERMES. As noted in Ref.~\cite{Aschenauer:2015rna}, it is important to perform the full integration in Eq.~\eqref{eq:def_multiplicities}, rather than evaluating the cross section at an average value for, in particular, $Q^2$. 

In order to study Hadron Mass Corrections in SIDIS, we will consider Nachtmann-type scaling variables defined by light-cone fractions of the photon's momentum $q$ and, respectively, the proton and hadron momentum. In the so-called ``$(p,q)$ frame'' \cite{Accardi:2009md}, in which $p$ and $q$ are collinear in 3-dimensional space and oriented along the $z-$direction, i.e., have zero transverse component ($\boldsymbol{p_T}=\boldsymbol{q_T}=\boldsymbol{0}$), one finds
\begin{align}
\label{eq:xi}
  \xi & \equiv -\frac{q^+}{p^+}
    = \frac{2 x_B}{1 + \sqrt{1 + 4 x_B^2 M^2/Q^2}} \\
\label{eq:zeta}
  \zeta_h & \equiv \frac{p_h^-}{q^-} = \frac{z_h}{2} \frac{\xi}{x_B}
    \left( 1 + \sqrt{1 - \frac{4 {x_B^2} M^2 m_{h}^2}{{z_h^2}\ Q^4}} 
    \right) \ ,
\end{align}
where $M$ is the nucleon target mass and $m_h$ is the mass of the detected hadron \cite{Guerrero:2015wha}. In the case of $\zeta_h$, note the interplay between initial and final state masses and Lorentz invariants that complicates the analysis. Had the data been binned in $z_e = - p\cdot p_h / q^2$, there would have been no mixing \cite{Guerrero:2015wha}. One can easily verify that in the Bjorken limit, where $M^2/Q^2 \to 0$ and $m_h^2/Q^2 \to 0$, one recovers the usual massless scaling variables $x_B$ and $z_h$.

\subsection{Collinear factorization with non-zero average parton virtualities}

At leading order (LO) in the strong coupling constant, one needs to calculate the diagram in Fig.~\ref{fig:SIDIS_handbag} left. The resulting hadronic tensor reads
\begin{equation}
  2MW^{\mu\nu} =
    \int d^4k\ d^4k'\
     \Tr \left[ \Phi_q(p,k)\, \gamma^\mu\, \Delta_q^h(k',p_h)\, \gamma^\nu
     \right] \, \delta^{(4)}(k + q - k') \, ,
  \label{eq:Wmunu}
\end{equation}
where $\Phi_q$ and $\Delta_q^h$ are quark-quark correlators associated with the quark distribution and fragmentation functions, respectively \cite{Collins:1989gx,Collins:1981uw,Bacchetta:2006tn,Mulders01}. For clarity of presentation, we are here considering only one quark flavor. In general, the right hand side would sport a charge-weighted sum over quark and antiquark flavors; details can be found in Ref.~\cite{Guerrero:2015wha}.

Obtaining a factorized expression for the hadronic tensor \eqref{eq:Wmunu} requires two steps. The first one is an expansion of the quark-quark correlators in inverse powers of the leading components of the parton momenta entering and exiting the hard scattering $H$ in Fig.~\ref{fig:SIDIS_handbag}, namely, $k^+$ and $k'^-$. In this paper, we limit our attention to the first order terms in this expansion, and write
$\Phi_q = k^+ \big[ \phi_2(k) \nbslash + {\cal O}(1/k^+) \big]$
and 
$\Delta_q = k'^- \big[ \delta_2(k') \nslash + {\cal O}(1/k'^-) \big]$.
The lowercase Greek letters indicate scalar functions of the momenta, and the unit light-cone plus-vector and minus-vector are denoted, respectively, by $n^\mu$ and $\bar n^\mu$. One then obtains
\begin{align}
  2MW^{\mu\nu} = \int d^4k\ d^4k'\ \phi_2(k) \delta_2(k')
    \Tr \left[ k^+\nbslash \, \gamma^\mu\, k^{\prime-}\nslash \,
    \gamma^\nu \right] \, \delta^{(4)}(k + q - k') + {\rm HT} \, .
  \label{eq:Wmunu-2}
\end{align}
In this expression, there is a clear separation between the partonic ``twist-2'' correlation functions $\phi_2$ and $\delta_2$ on the one hand, and  on the other hand the dynamics and overall kinematics of the hard scattering process (namely the trace and $\delta$-function).
Note that the neglected pieces are not forgotten, but in fact contribute to restore gauge invariance in ``higher-twist'' (HT) diagrams that include additional parton exchanges between the hard scattering and the non-perturbative blobs in Fig.~\ref{fig:SIDIS_handbag} \cite{Qiu:1988dn,Ellis:1982cd,Bacchetta:2006tn}.

The second step consists in approximating the incoming and outgoing partonic momenta appearing in the four-momentum conservation $\delta$-function, namely $k \approx \widetilde{k}$ and $k'\approx \widetilde{k'}$.  It is important to remark that this is the only place where we perform an approximation rather than a controlled expansion. Contrastingly, the trace part is expanded in inverse powers of the plus and minus light-cone momenta; one could then improve on this approximation by retaining higher order terms, and considering in addition diagrams involving multi-parton correlators.  After contraction with the leptonic tensor, the result would be a ``twist'' expansion of the SIDIS cross section in powers of $\Lambda/Q$, where $\Lambda$ is a scale quantifying the strength of parton-parton correlations inside the proton target and the detected hadron. In this paper, however, we only consider terms of order $(\Lambda/Q)^0$.

In collinear factorization, one chooses $\widetilde{k}$ and $\widetilde{k'}$ to be collinear in 3D space to the momentum of the target nucleon and the detected hadron, respectively. In the $(p,q)$ frame, these read
\begin{align}
  k^{\mu} \approx \widetilde k^{\mu}
    & = \Big(xp^+,\frac{v\,^2}{2xp^+},\bm{0}_T\Big) \label{eq:k} \\
  k'^{\mu} \approx \widetilde k'^{\mu}
    & = \Bigg(\frac{\vpsq +
      (\bm{p_{h\perp}}/z)^2}{2p_h^-/z},
      \frac{p_h^-}{z},\frac{\bm{p_{h\perp}}}{z}\Bigg) \ .
\label{eq:kp}    
\end{align}
As a consequence, the hadronic tensor turns out to depend only on the 1-dimensional, ``collinear'' Parton Distribution Function $q(x) = \int dk^-d^2k_T \phi_2(k)$, and Fragmentation Function $D_q(z) = (z/2) \int dk'^+d^2k'_T \delta_2(k)$, where $x \equiv k^+/p^+$ and $z \equiv p_h^-/k'^-$.

At variance with the conventional treatment, we consider generic  ``average virtualities'' $v\,^2 \approx \langle k_\mu k^\mu\rangle$ and $\vpsq \approx \langle k'_\mu k^{\prime \mu}\rangle$ for the incoming and outgoing partons, whose values will be fixed later rather than put to 0. It is indeed clear from Fig.~\ref{fig:SIDIS_handbag} that $k'_\mu k^{\prime \mu} \geq m_h^2$, and that this bound cannot be {\it a priori} neglected at the kinematics we are interested in. In general, the average virtualities of the quarks entering the diagram in Fig.~\ref{fig:SIDIS_handbag} are determined by the dynamics of the scattering and hadronization processes, see for example Refs.~\cite{Collins:2007ph,Moffat:2017sha}, and can be in principle different for the scattering and scattered quarks. 
It is also important to to keep the quark's current mass $m_q$ and virtuality $v$ or $v'$ conceptually separated. It is only when a quark line is cut, {\it i.e.}, when the scattered quark appears in the final state, that $k'_\mu k^{\prime \mu} = m_q^2$; this is clearly not the case for either quark in the handbag diagram of Fig.~\ref{fig:SIDIS_handbag}. Furthermore, it is $m_q$, rather than the virtualities $v$ or $\vp$ (as claimed in Ref.~\cite{Christova:2016hgd}), that appears, via Dirac's equation, in the so-called ``equations of motion relations'' essential to the treatment of HT terms \cite{Bacchetta:2006tn}.

Finally, the SIDIS hadronic tensor factorizes into a convolution of a quark PDF, a quark FF, and a hard scattering tensor ${\cal H}^{\mu\nu}$ as \cite{Guerrero:2015wha}
\begin{align}
  2MW^{\mu \nu}
  & = \sum_q e_q^2 \int \frac{dx}{x} \frac{dz}{z} \,
  q(x) \, {\cal H}^{\mu\nu}(x,z) \, D_q(z) + {\rm HT} \ ,
\label{eq:Wmunu_HMC}
\end{align}
where
\begin{align}
  {\cal H}^{\mu\nu}(x,z)
    & = \frac{1}{2z}\Tr\big[ \kslash_0 \gamma^\mu \kpslash_0 \gamma^\nu \big] \,
      \delta\Big(k_0^+ + q^+ - \frac{\vpsq}{2k_0^{\prime-}} \Big) \,
      \delta\Big(\frac{v^2}{2k_0^+} + q^- - k_0^{\prime-} \Big) \,
      \delta^{(2)}(\bm{k'_{0T}}) \ .
\label{eq:partonic-H}
\end{align}
For ease of interpretation and discussion, in this formula we explicitly separated the virtualities $v$ and $v'$ from  the ``massless'' partonic momenta $k_0$ and $k'_0$ defined as
\begin{align}
  k_0^\mu & \equiv \widetilde k^\mu|_{v=0} = (xp^+,0,\bm{0_T}) \\
  k_0^{\prime\mu} & \equiv \widetilde k^\mu|_{\vp=0} = (0,p_h^-/z,\bm{p_{hT}}/z) \ .
\end{align}
In Eq.~\eqref{eq:Wmunu_HMC}, we reinstated the sum over quark flavors and neglected terms of twist higher than 2 in the twist expansion; a detailed discussion of factorization at twist 3 and twist 4 in inclusive and semi-inclusive DIS can be found in Refs.~\cite{Qiu:1988dn,Ellis:1982cd,Bacchetta:2006tn,Wei:2016far}. 
Let us remark that, {\it as a result} of our calculation and factorization scheme, the trace term appearing in Eq.~\eqref{eq:partonic-H} can be {\it interpreted} as the matrix element squared for the scattering of a virtual photon with a parton of momentum $k_0^\mu$ and {\it current} mass $m_q=0$; however, when the average virtualities $v$ and $\vp$ have non-zero values,
the $\delta$-functions impose different values of $x$ and $z$ than if this was an actual physical process. 

It is interesting to note that, if one does choose $v=\vp=0$, the hard scattering tensor can be rewritten as
\begin{align}
 {\cal H}^{\mu\nu}|_{v=\vp=0}
    & \propto \Tr\big[ \kslash_0 \gamma^\mu \kpslash_0 \gamma^\nu \big] \,
      \delta^{(4)} \big(k_0 + q - k'_0 \big) \ .  
\label{eq:partonic-H_PM}
\end{align}
Then the whole SIDIS hadronic tensor can be {\it interpreted} in terms of the parton model scattering and fragmentation of a fictitious free quark of zero mass collinear with the target nucleon and the produced hadron. This model was proposed by Feynman as a heuristically well motivated approximation to the full QCD process in the ``infinite momentum'' $p^+\sim Q$ frame \cite{Feynman:1973xc}; however, at sub-asymptotic values of $Q$, such as those investigated here, the resulting approximation may not be optimal. In the original parton model, the masses of the target and of the detected hadron are neglected. Our Eq.~\eqref{eq:partonic-H_PM}, instead, supplements the parton model with mass corrections in a way that was already proposed by Albino et al. in Ref.~\cite{Albino:2008fy} and shown to provide improved fits of experimental data.

If, however, one chooses $v\neq0$ or $\vp\neq0$, the $\delta$-functions in Eq.~\eqref{eq:partonic-H} cannot be interpreted as expressing four momentum conservation of the fictitious free quark as it scatters on the virtual photon; therefore, the hard scattering tensor cannot be given a parton model interpretation. This lack of intuitive interpretability is not to be considered a show stopper: on the contrary, the hard scattering tensor \eqref{eq:partonic-H} satisfies by inspection the Ward identity $q_\mu {\cal H}^{\mu\nu}=0$, and therefore the hadronic tensor \eqref{eq:Wmunu_HMC} is a legitimate, gauge invariant approximation of the full scattering diagram in Fig.~\ref{fig:SIDIS_handbag}.
In fact, as argued in Refs~\cite{Accardi:2009md,Guerrero:2015wha} and discussed next, $\vp\neq0$ is a necessary condition to respect 4-momentum conservation in the SIDIS process. Thus, our Equations~\eqref{eq:Wmunu_HMC}-\eqref{eq:partonic-H} provide the means to go beyond the parton model, and to implement this kinematic requirement in a gauge invariant way in the collinear factorization framework.
 
More in general, the 2-steps procedure discussed above defines an ``approximator'' of the hard scattering which is analogous to that introduced by Collins, Rogers and Stasto in Ref.~\cite{Collins:2007ph}. Compared to that one, our approximator takes into account kinematical hadron mass effects, and furthermore allows one to define fully integrated collinear PDFs and FFs instead of the totally unintegrated ones considered in the mentioned reference. For a full proof of factorization, one would furthermore need to verify that this scheme extends at least to NLO, and that our approximator allows one to use Ward identities to factor out and resum longitudinal gluons into the Wilson lines needed to ensure gauge invariance in the operator definition of the PDFs and FFs. The successful phenomenological approach of our LO scheme, to be discussed in detail in Section~\ref{sec:multiplicities}, justifies future efforts in this direction. 

\subsection{Choice of virtualities}

Upon integration over the delta functions, one obtains
\begin{subequations}
\begin{eqnarray}
  \frac{x}{\xi} & = & 1 + \frac{z}{\zeta_h}\frac{\vpsq}{Q^2}
                      \label{eq:hard_scattering_I} \\
  \frac{\zeta_h}{z} & = & 1 + \frac{\xi}{x}\frac{v\,^2}{Q^2} \ ,
                      \label{eq:hard_scattering_II}
\end{eqnarray}
\end{subequations}
and, clearly, in the Bjorken limit, one recovers $x \approx x_B$ and $z \approx z_h$. To proceed further, it is necessary to specify a choice for the average virtualities $v$ and $\vp$. For this purpose, we minimally require that the approximated, internal $\widetilde k$ and $\widetilde k'$ momenta respect four-momentum and baryon number conservation in the factorized diagram, or in other words that the ``internal'' (approximated) collinear kinematics at parton level matches the ``external'', hadron-level kinematics. The limits that this requirement places on the possible values of $v$ and $\vp$ have been derived in detail in Refs.~\cite{Accardi:2009md,Guerrero:2015wha}. Here we only recall the main results, and defer to Appendix~\ref{app:baryon_number} a subtler point regarding the treatment of baryon number conservation that was not sufficiently explained in those papers.

Due to the interpretation of the trace term in the hard scattering tensor ${\cal H}^{\mu\nu}$ as due to a ``massless'' quark of momentum $\tilde k_0$ scattering on a virtual photon, it is desirable to choose 
\begin{align}
  v^2 = 0 \ .
\label{eq:v=0}
\end{align}
More importantly, if we applied this formalism to semi-inclusive hadron production in electron-positron scattering events, a value $v^2=0$ would be imposed by the cut on the non-fragmenting (light) quark line in the leading order diagram. Thus, Eq.~\eqref{eq:v=0} is in fact a necessary condition for the proposed HMC formalism to be universal. Fortunately, a zero value for $v$ is also compatible with the external kinematics \cite{Accardi:2009md,Guerrero:2015wha}. As a result, Eq.~\eqref{eq:hard_scattering_II} requires $z=\zeta_h$.

The fragmentation of the scattered parton into a massive hadron, instead, requires a non vanishing virtuality $\vpsq$. More precisely, by requiring four momentum conservation in the right hand side diagram of Fig.~\ref{fig:SIDIS_handbag} ({\it i.e.}, by matching the internal approximated partonic kinematic with the external hadronic kinematics of the fragmentation process), one finds that a parton of light-cone momentum fraction $z$ needs an average virtuality $\vpsq \geq m_h^2/z$ to fragment into a hadron of mass $m_h^2$. Then, compatibly with the LO constraint on $z$ just discussed, we choose
\begin{align}
  \vpsq = m_h^2/\zeta_h \ .
\label{eq:}
\end{align}
Inserting these virtualities into Eqs.~\eqref{eq:hard_scattering_I}-\eqref{eq:hard_scattering_II}, one finds
\begin{eqnarray}
x &=& \xi_h \equiv \xi \Big(1 + \frac{m_h^2}{\zeta_h Q^2} \Big)
\label{eq:xi_h} \\
z &=& \zeta_h   \ ,
\label{eq:zeta_h}
\end{eqnarray}
with the Nachtmann-type scaling variables $\xi$ and $\zeta_h$ defined in Eqs.~\eqref{eq:xi}-\eqref{eq:zeta}.

Note that $\xi_h$ is reminiscent of the $\chi=x_B(1+4m_Q^2/Q^2)$ scaling variable in the ACOT-$\chi$ treatment of heavy quarks in DIS \cite{Tung:2001mv,Nadolsky:2009ge}, where an unobserved open heavy flavor of mass $2m_Q$ is produced in the final state, much like the detected hadron of mass $m_h$ and momentum fraction $z_h$ discussed in this paper. Further exploration of this similarity is left for future work.

\subsection{Factorized hadron multiplicities}

Collecting the above results, contracting the hadronic tensor with the leptonic tensor, and accounting for mass corrections also in the inclusive cross section appearing in the denominator \cite{Accardi:2008ne}, one can see that, at finite $Q^2$ values, the LO hadron multiplicity integrand in Eq.~\eqref{eq:def_multiplicities} factorizes in terms of products of quark PDFs and FFs, $D_q^h$, but evaluated at the scaling variables $\xi_h$ and $\zeta_h$ just derived, and that 
\begin{equation}
M^{h}(x_B^{\textrm exp}) = \frac{\sum_q e_q^2\, \int_{\textrm exp} dx_B dQ^2 \int_{0.2}^{0.8(0.85)}\, dz_h\, J_h \,  q(\xi_h,Q^2)\, D_q^h(\zeta_h,Q^2)}{\sum_q e_q^2\, \int_{\textrm exp} dx_B dQ^2 \, q(\xi,Q^2)} \ ,
\label{eq:finite_Q2_multiplicities}
\end{equation}
where $J_h = d\zeta_h/dz_h$ is a scale-dependent Jacobian factor \cite{Guerrero:2015wha}. This expression is gauge invariant and incorporates the kinematic requirement for the scattered  quark to have a non-zero virtuality in order to fragment into a hadron of non-zero mass $m_h$ and invariant momentum fraction $z_h$. Note that in the Bjorken limit all masses become negligible ($m^2/Q^2 \to 0$) and one recovers the usual ``massless'' $M_h^{(0)}$ multiplicity,
\begin{equation}
M^{h (0)}(x_B^{\textrm exp}) = \frac{\sum_q e_q^2\, \int_{\textrm exp} dx_B dQ^2 \, q(x_B, Q^2)\,\int_{0.2}^{0.8(0.85)} dz_h D_q^h(z_h, Q^2)}{\sum_q e_q^2 \int_{\textrm exp} dx\,dQ^2\,  q(x_B, Q^2)} \ ,
\label{eq:partonlevel_multiplicities}
\end{equation}
with its usual parton model interpretation.

The arguments leading to the factorized formula \eqref{eq:finite_Q2_multiplicities} and the proof of its gauge invariance were already laid out in Refs.~\cite{Accardi:2009md,Guerrero:2015wha}, although in a way that seems to have originated some misunderstanding and confusion in the literature \cite{Christova:2016hgd}. It is our hope that the present discussion will dispel the doubts raised there on the validity of our treatment of hadron mass corrections.

\section{Numerical results for kaon multiplicities}
\label{sec:multiplicities}

The HERMES and COMPASS data on integrated kaon multiplicities \cite{Airapetian:2012ki,Seder:2015sdw,Adolph:2016bwc} appear to be incompatible with each other, a well known and hotly debated fact \cite{Aschenauer:2015rna,Stolarski:2014jka,Leader:2015hna}. While most of the discussion has centered on kinematic and binning issues, here we argue that HMCs may also play an essential role due to the relatively low average values of $Q^2$ dominating the HERMES and COMPASS measurements.

\subsection{Data over theory ratios}

One way to compare HERMES multiplicities to COMPASS multiplicities is through the ratio between experimental data and theory calculations, in which both the difference in kinematic cuts and  $Q^2$ evolution between the two experiments are canceled, having been included in the theory calculation Eq.~\eqref{eq:def_multiplicities}: with a perfect theory (and in the absence of unaccounted for experimental systematics) the ratio should be equal to 1 within statistical fluctuations.

\begin{figure}[htb]
  \centering
  \includegraphics[width=18cm]{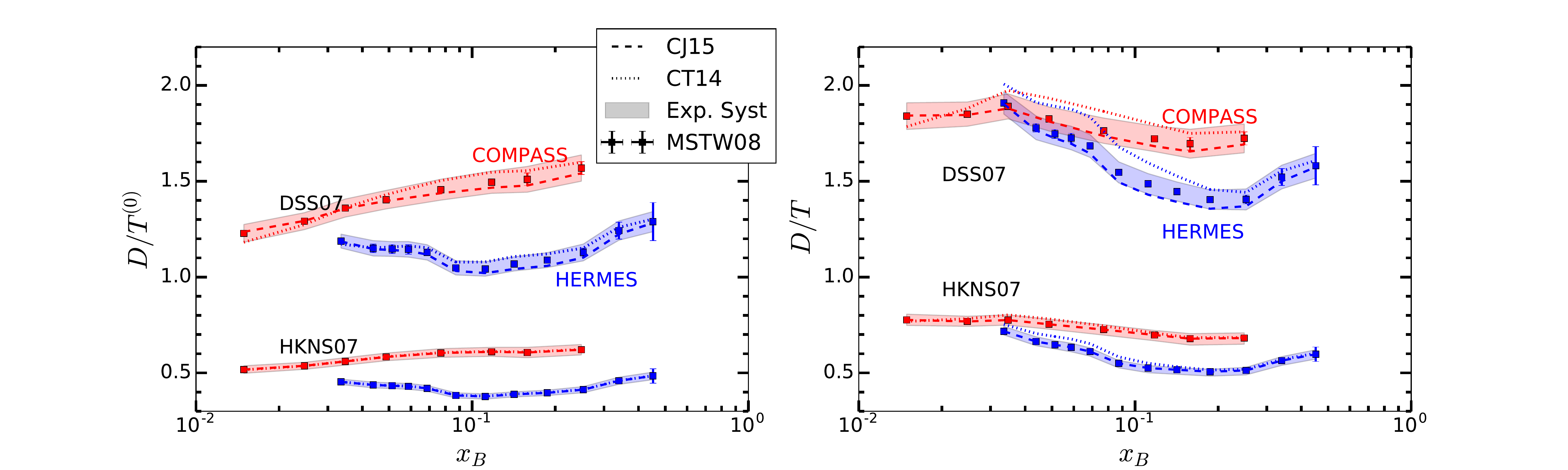}	
  \caption{Ratio of experimental data ($D$) over theory predictions for $K^++K^-$ SIDIS multiplicities as a function of $x_B$ for the HERMES (blue line) and COMPASS (red line) experiments on deuterium targets. The left plot shows the ratio using the massless theory ($T^{(0)}$), while in the right plot the finite-$Q^2$ theory ($T$) is used.}
  \label{data_over_theory}.
\end{figure}

In Fig.~\ref{data_over_theory}, we can observe the data over theory $D/T$ ratio for different sets of PDFs \cite{Martin:2009iq,Accardi:2016qay,Dulat:2015mca} and FFs \cite{deFlorian:2007aj,Hirai:2007cx}.
The effect of HMCs can be observed comparing the ratio calculated using the massless theory (left panel) and the theory with HMCs (right panel). There clearly is a large FF systematics due to the poor knowledge we have of kaon fragmentation functions, but this amounts largely to an overall multiplicative factor; the PDF systematics is definitely smaller.

In the massless ratios, even looking at only one given FF set, one can notice a difference in size, as well as shape, of the HERMES and COMPASS $D/T$ ratios. Using HMCs the size discrepancy between the two experiments is reduced and the ratio is flatter for both sets of data. In particular the COMPASS ratio is rather flat over the whole $x_B$, while HERMES still persists having  a downward slope and a concave shape.

\subsection{Multiplicities in a massless world}
\label{massless_world}

A more direct data-to-data comparison of HERMES and COMPASS results, that also reduces the effect of the FF and PDF systematics, can be obtained by defining ``theoretical correction ratios'' that produce (approximate) massless parton multiplicities at a common beam energy.
This method also allows one to interpret the corrected data at face value using parton model formulas such as our $M_h^{(0)}$ in Eq.~(\ref{eq:partonlevel_multiplicities}), or Eq.~(2) of Ref.~\cite{Airapetian:2013zaw}.

The first step in the calculation consists in removing the mass effects from the original data using the ``HMC ratio''
\begin{equation}
  R^h_{HMC} = \frac{M^{h(0)}}{M^h} \ ,
  \label{eq:R_HMC}
\end{equation}
where $M^{h(0)}$ is the massless hadron multiplicity calculated theoretically using Eq.~(\ref{eq:partonlevel_multiplicities}) and $M^{h}$ is the finite $Q^2$ multiplicity from Eq.~(\ref{eq:finite_Q2_multiplicities}). Using this, the product $M_{exp}^h\times R^h_{HMC}$ can be interpreted as a ``massless'' experimental multiplicity. In other words, this is the multiplicity that one would expect to measure in a world where nucleons and kaons are massless.

Next, we address evolution effects, {\it i.e.}, the difference in the $Q^2$ reach of each $x_B$ bin of HERMES and COMPASS. For this, we choose the COMPASS kinematics to be the one at which we want to compare the data. Then, we bring HERMES data to COMPASS energies through an evolution ratio that we define as:
\begin{equation}
R^{H \rightarrow C}_{evo} = \frac{M^{h(0)}(x_B^{HERMES})\Bigr\rvert_{\textrm{COMPASS cuts}}}
{M^{h(0)}(x_B^{HERMES})\Bigr\rvert_{\textrm{HERMES cuts}}} \ .
\label{eq:R_evo}
\end{equation}
Here, the numerator is the massless multiplicity calculated integrating over each one of the HERMES $x_B$ bins, but using the kinematic acceptance of the COMPASS experiment; namely, we integrate over the black hatched vertical stripes in the $(x_B,Q^2)$ phase space shown in the right panel of Fig.~\ref{fig:H_C_PS}. 
The denominator is the massless multiplicity integrated using the original HERMES kinematic cuts (blue vertical stripes in the right panel of Fig.~\ref{fig:H_C_PS}).
As a result, multiplying the massless HERMES multiplicity found in the previous step by this ratio, we are effectively ``evolving'' HERMES results to COMPASS energy and spectrometer.

\begin{figure}[t]
  \centering
  \includegraphics[width=8cm]{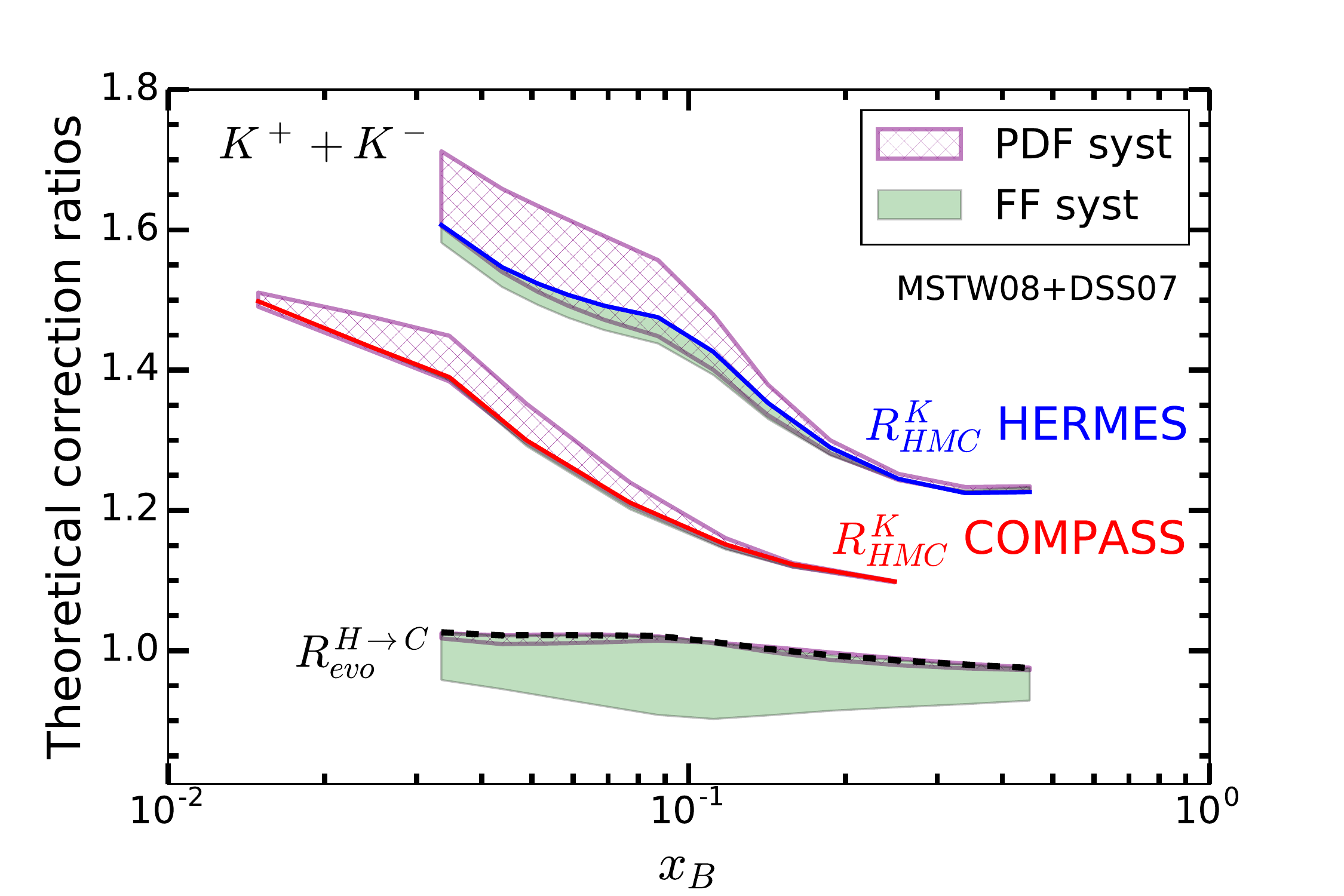}		
  \caption{Theoretical correction ratios for charged $K^+ + K^-$ multiplicity as a function of $x_B$ for Mass Corrections at HERMES (blue line), COMPASS (red line), and HERMES-to-COMPASS evolution (black dashed line). PDF and FF systematic uncertainties are plotted, respectively, as purple hashed and green shaded band. }
  \label{correction_ratios}
\end{figure}

\begin{figure}[t]
	\centering
	\includegraphics[width=8cm]{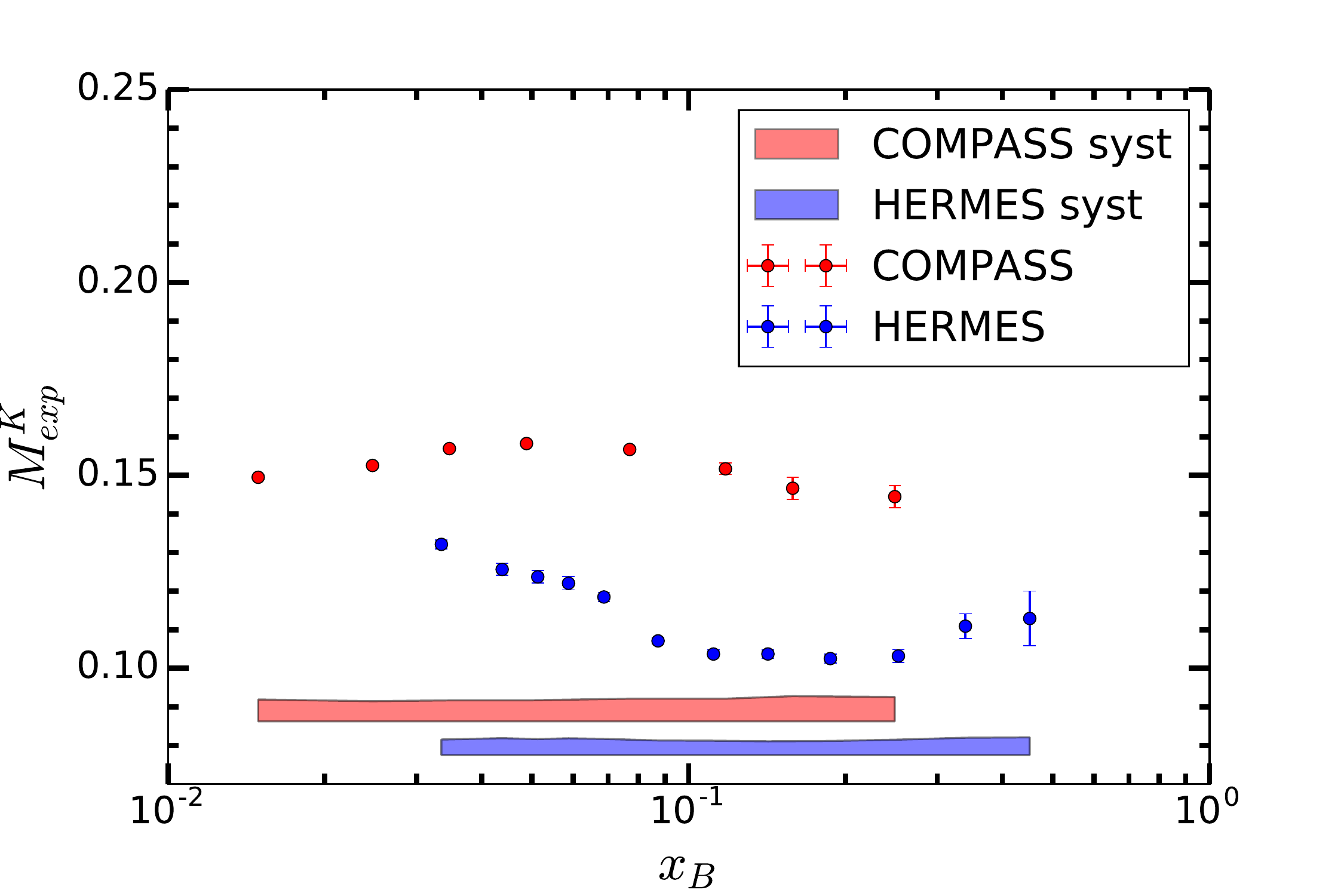}	
	\includegraphics[width=8cm]{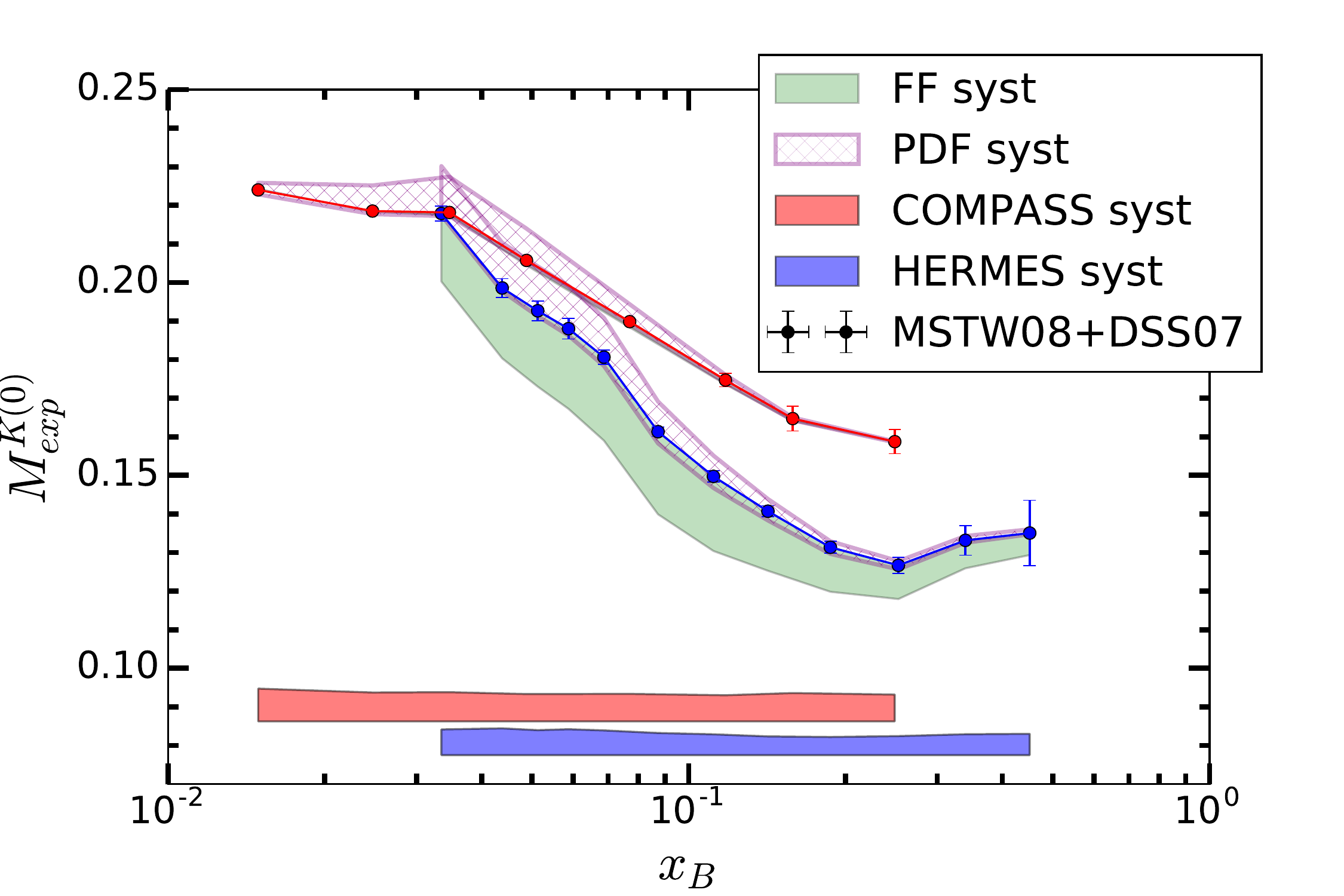}		
	\caption{Right: Experimental data for integrated kaon Multiplicities ($K^+ + K^-$). Left: Parton level multiplicities after applying the theoretical correction ratios given by Eq.~(\ref{eq:data_sum_corrected}) to the data shown on the right. The green FF systematics band for COMPASS is very small compared to the HERMES case, and almost invisible in the plot. This is due in part to the larger $Q^2$ value in each bin, and in part to the additional systematics in the HERMES band introduced by the evolution ratio in Eq.~(\ref{eq:HERMES_corrected}).}	
	\label{corrected_data_sum}
\end{figure}

Finally, the massless and evolved experimental multiplicities can be defined by multiplying the original data by the appropriate correction ratios; in our case,
\begin{subequations}
	\begin{eqnarray}	
	M_{exp}^{h(0)} & \equiv \ M_{exp}^h \times R^h_{HMC}\hspace{2.07cm}   &\textrm{(for COMPASS)}
	\label{eq:COMPASS_corrected} \\ 
	M_{exp}^{h(0)} & \equiv \ M_{exp}^h \times R^h_{HMC} \times R_{evo}^{H \rightarrow C} \hspace{0.74cm}  & \textrm{(for HERMES)}.
	\label{eq:HERMES_corrected}
	\end{eqnarray}
	\label{eq:data_sum_corrected}%
\end{subequations}
The correction ratios were evaluated numerically and plotted in Fig.~\ref{correction_ratios} and we find, as expected, that these are relatively stable with respect to the choice of FFs, because the FF systematics shown in Fig.~\ref{data_over_theory} is canceled in the ratios defined in Eqs.~(\ref{eq:R_HMC}) and ~(\ref{eq:R_evo}). The PDF systematics is also small. Furthermore, hadron mass effects are dominant compared to evolution effects, that are rather small. For COMPASS, the HMC corrections are smaller than at HERMES because the $Q^2$ accessed at COMPASS is higher at a given $x_B$ than at HERMES due to the higher beam energy. The PDF and FF systematic uncertainties are calculated by varying these among the fits listed in Fig.~\ref{data_over_theory}, and are typically smaller at COMPASS due to the higher $Q^2$ reach. The green FF systematic band for the COMPASS $R^{K}_{HMC}$ is very small compared to the HERMES case, and almost invisible in the plot. The purple PDF systematic band for $R_{evo}^{H \rightarrow C}$ is very small compared to the FF green band.

In Fig.~\ref{corrected_data_sum}, we plot the experimental $K^+ + K^-$ multiplicity data $M_{exp}^K$ on the left and the ``massless'' multiplicities $M_{exp}^{K (0)}$ on the right using Eqs.~(\ref{eq:COMPASS_corrected})-(\ref{eq:HERMES_corrected}).
In the $D/T$ ratios, HMCs were included in the theoretical calculations; here, instead, HMCs are ``removed'' from data. After furthermore evolving HERMES data to COMPASS energy (which was automatically achieved in the $D/T$ ratios), the discrepancy in size between HERMES and COMPASS is also largely reduced. Moreover, the corrected data, which can be interpreted directly in terms of parton model formulas, now show for both experiments a negative slope in $x_B$ that agrees much better with the $(1-x)^\beta$ power law behavior of any PDF, including the s-quark. Clearly the slopes and shapes in $x_B$ of the HERMES and COMPASS data do not match yet, which indicates that corrections other than HMCs, or unquantified systematic uncertainties, are at play.

\section{Numerical results for Kaon multiplicity ratios}
\label{sec:ratios}

Another interesting observable is the $K^+/K^-$ multiplicity ratio, because one can expect the systematic and theoretical uncertainties in each experiment, as well as $Q^2$ evolution effects, to largely cancel between numerator and denominator. However, one may still expect some residual mass effect because of the different slopes in $z$ of the $K^+$ and $K^-$ FFs. 

\begin{figure}[t]
  \centering
  \includegraphics[width=17cm]{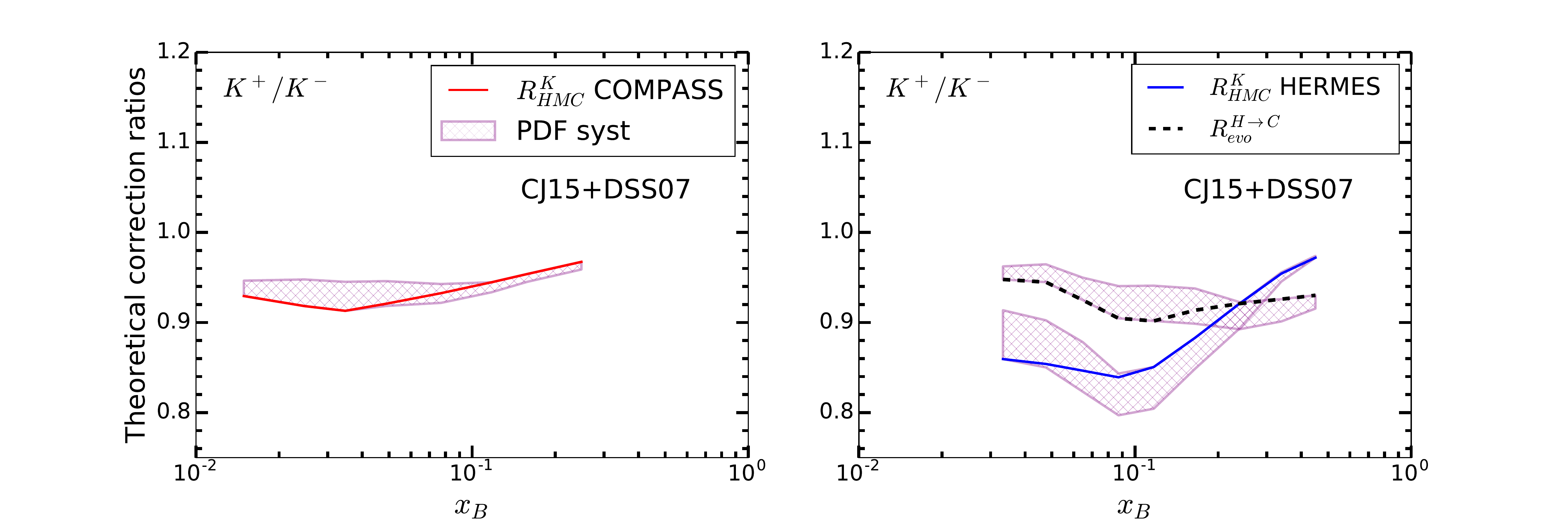}		
  \caption{Theoretical correction ratios for charged $K^+/ K^-$ multiplicity ratio as a function of $x_B$ for mass corrections at COMPASS (red line, left panel) and HERMES (blue line in the right panel), and for HERMES-to-COMPASS evolution (black dashed line, right panel).
  PDF systematic errors are plotted as a purple hashed band. Note the difference in vertical scale compared to Fig.~\ref{correction_ratios}.}
  \label{correction_ratios_p_m}
\end{figure}

The theoretical correction ratios for the $K^+/K^-$ multiplicity ratio are plotted in Fig.~\ref{correction_ratios_p_m}. As expected, the corrections are smaller than in the  $K^+ + K^-$ multiplicity sum. The HMCs are non negligible (up to $-15 \%$ for HERMES and $-10 \%$ for COMPASS) and of the same order of the HERMES to COMPASS evolution effects. (The FF systematics has not been evaluated because the HKNS fit cannot extract reliable charge separated fragmentation functions.)

\begin{figure}[t]
	\centering	
	\includegraphics[width=8cm]{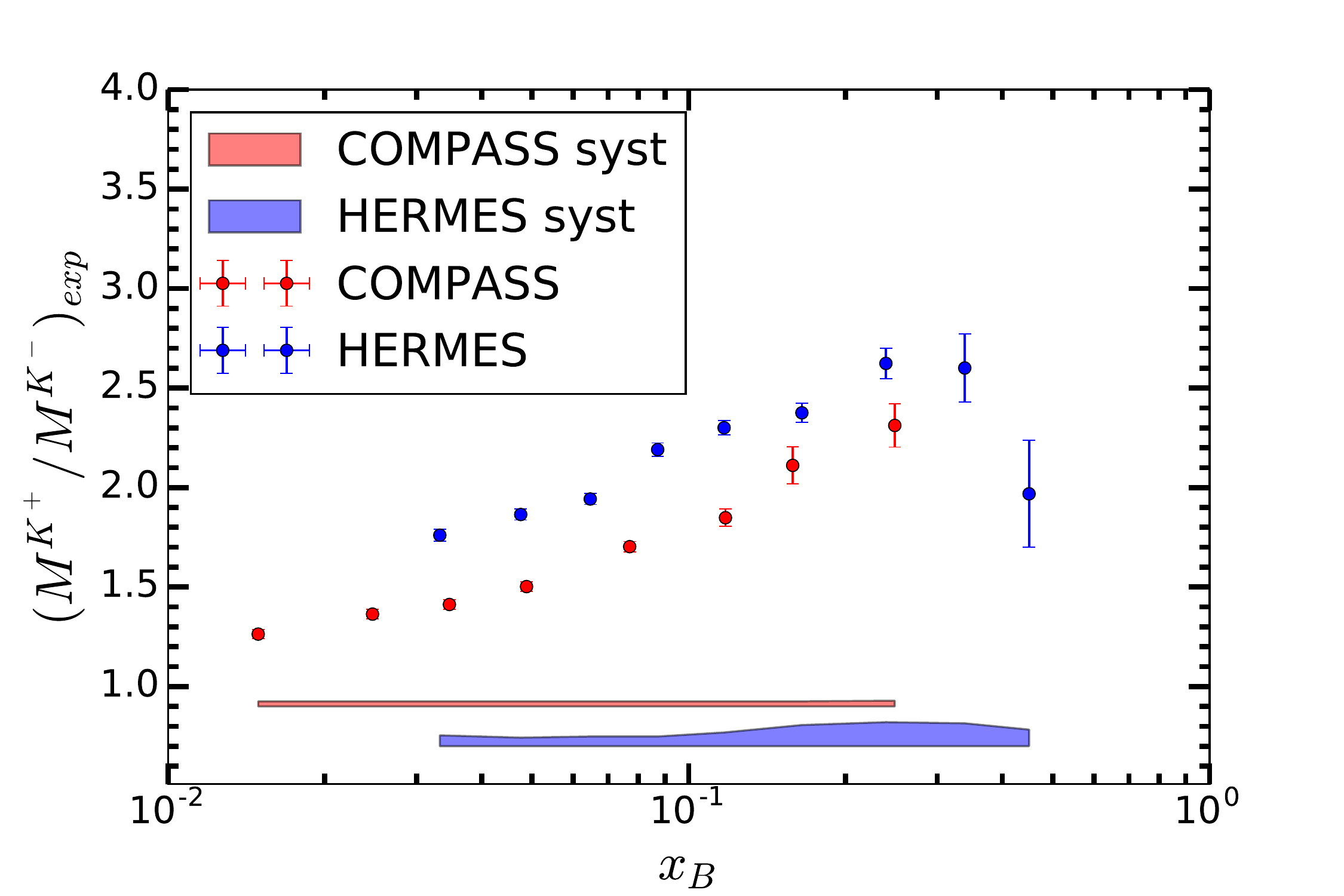}	
	\includegraphics[width=8cm]{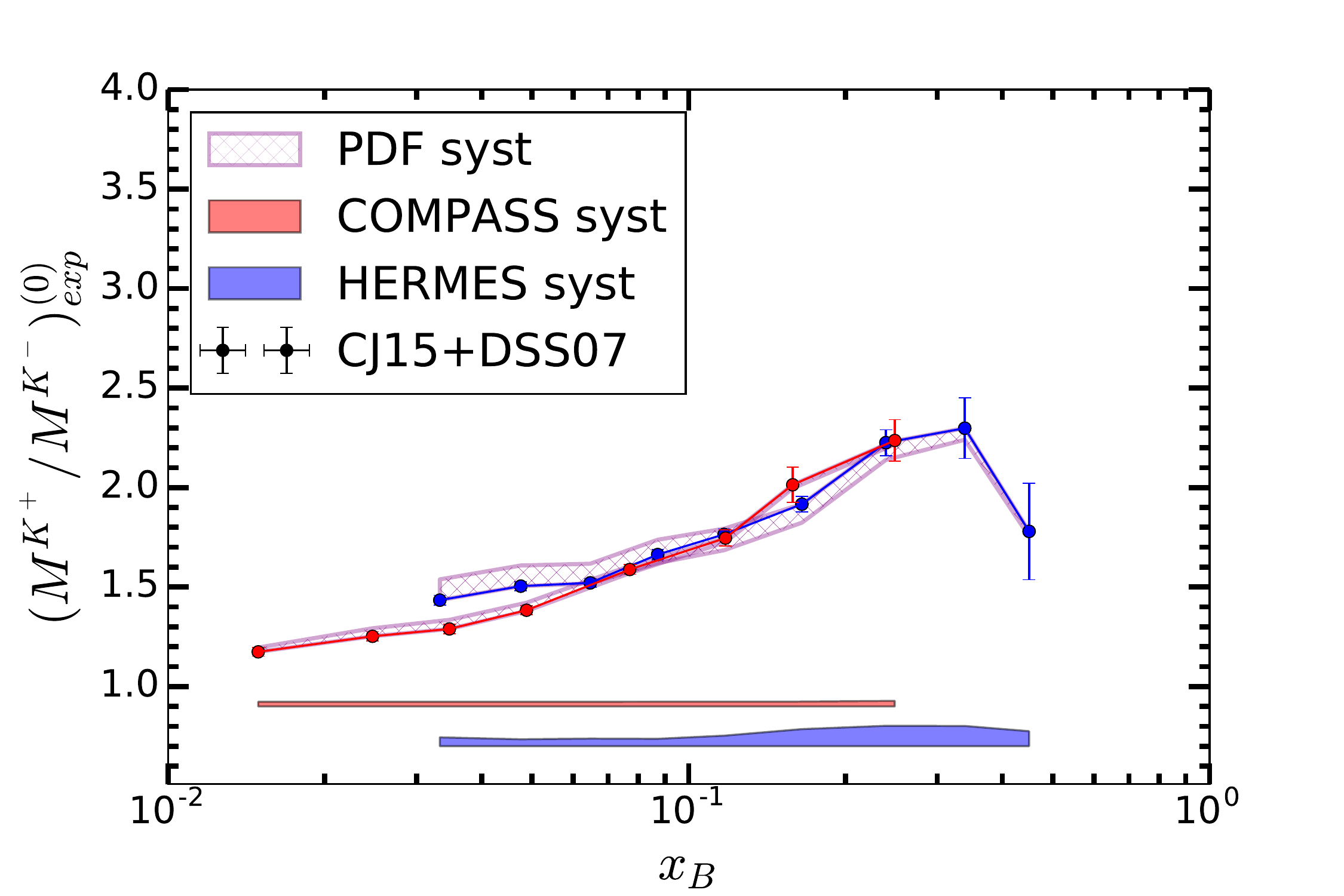}	
	\caption{Right: Experimental data for integrated kaon Multiplicities ($K^+ / K^-$). Left: Parton level multiplicities after applying the theoretical correction ratios given by Eq.~(\ref{eq:data_sum_corrected}) to the data shown on the right.}
	\label{corrected_data_ratio}
\end{figure}

The original and ``massless'' data for both HERMES and COMPASS experiments are then plotted in Fig.~\ref{corrected_data_ratio}. In this case, the slopes are compatible already in the original data, which shows that much of the systematics difference between the two experiments is not irreducible, but affects only the charged $K^++K^-$ multiplicities. However, the discrepancy in size persists. After removing the mass effects and compensating for evolution, the ``massless'' kaon ratios become fully compatible between the two experiments. A possible exception is the last HERMES $x_B$ bin, that shows a sharp change in slope as it also happens for the case of the summed $K^+ + K^-$, but lies just outside the COMPASS range. In the charged multiplicity case, this could be partly attributed to nuclear binding and Fermi motion effects in the Deuteron target. However, nuclear effects should largely cancel out in the $K^+/K^-$ ratio, and the origin of the slope change (which is however marginally compatible with the rest of the data within systematic and statistical uncertainties) remains to be understood.

\section{Other $\bm{Q^2}$-dependent corrections}
\label{sec:otherQ2}

The HMC calculations presented in the previous sections have been performed at leading twist and leading order in $\alpha_s$ accuracy. As these do not seem to exhaust the sources of difference between HERMES and COMPASS integrated multiplicity ratios (although they can potentially explain just by themselves the difference in the kaon ratio measurements) it is worthwhile commenting on other $Q^2$-dependent corrections.

Higher-Twist contributions in unpolarized scattering scale as $\Lambda^2/Q^2$, where $\Lambda$ is a dynamical non-perturbative scale that quantifies quark-gluon correlations inside the nucleons. Since this is the same kind of scaling exhibited by HMCs, that are however kinematic in origin and scale as $m_K^2/Q^2$, one may wonder if HT corrections might explain the residual difference in kaon multiplicities. This is certainly possible, although quantifying those corrections is outside the scope of this article. Here, we just note that the HT phenomenology in inclusive DIS is well developed \cite{Accardi:2016qay,Virchaux:1991jc,Alekhin:2003qq,Blumlein:2008kz}, while we are not aware of similar studies for SIDIS.

Likewise, one may want to consider NLO corrections, that, however, depend only logarithmically on $Q^2$. These may therefore slightly tilt the data/theory ratios for massless multiplicities, but not necessarily close the remaining gap between the HERMES and COMPASS data, as also suggested by the calculations presented in Ref. \cite{WenKaoDIS2018}. This will be explored in a forthcoming paper \cite{AGinprep}.

\section{Discussion and concluding remarks}
\label{sec:conclusions}

In this paper, we have argued that Hadron Mass Corrections of order ${\mathcal{O}(m^2/Q^2)}$ in SIDIS are non negligible for kaon production at HERMES and COMPASS, where integrated multiplicities have an average $Q$ ranging from $1$ to $4$ GeV, quite comparable to the kaon mass. These corrections can be captured in a gauge-invariant way at leading twist by new massive scaling variables that incorporate the need for the struck quark to be sufficiently off the free-particle mass shell in order to fragment into a massive hadron.
At leading order in the coupling constant, the leading-twist cross section still factorizes into a product of PDFs and FFs, but is evaluated at the Nachtmann variable $\xi_h$ of Eq.~(\ref{eq:xi_h}) and the fragmentation scaling variable $\zeta_h$ of Eq.~\eqref{eq:zeta}, respectively. 

After accounting for HMCs in this way, we found that the discrepancy between the integrated kaon multiplicities measured by the HERMES and COMPASS collaborations is reduced.

For the charge-summed $K^+ + K^-$ multiplicity there are still some differences in slope and shape that need to be investigated. From a theoretical side, one would certainly need to evaluate the effects of Higher-Twist contributions, while NLO corrections, that scale logarithmically in $Q^2$, do not seem likely to close the gap remaining between the two experimental measurements. 

In the case of the $K^+/K^-$ multiplicity ratio, where much of the theoretical and experimental systematics can be expected to cancel, the slopes were already similar in the published data, and HMCs can fully reconcile the remaining discrepancy in size. The only possible exception are the last two $x_B$ bins of the HERMES measurement, that however lie just outside the reach of the COMPASS experiment. It would therefore be interesting to repeat these measurements at the 12 GeV Jefferson Lab upgrade (JLab 12), where a higher $x_B$ range could be covered at $Q^2$ values comparable to the average HERMES $Q^2$, but retaining nonetheless a considerable overlap in $x_B$ with both HERMES and COMPASS. Likewise, measuring pion multiplicities at JLab 12 would allow one to investigate the large difference in that overlap region between existing measurements at Jefferson Lab and HERMES noted in Ref.~\cite{Adolph:2016bga}, but with an intermediate energy beam. 

The nearly perfect agreement in the overlap region of the kaon multiplicity ratios after HMCs are taken into account is a strong indication that the remaining differences in the charged kaon multiplicities are of systematic origin -- whether theoretical or experimental remains to be ascertained. This conclusion is strengthened by observing that in the case of the much lighter (and essentially HMC-free) pion, the ratios measured by the two experiments also agree despite displaying strong differences in the charged multiplicity data.

As an outlook, we would like to include deuteron nuclear corrections in our analysis to see if this may explain the large $x_B$ behavior of the HERMES data. More importantly, however, we need to prove that factorization extends to NLO in perturbation theory when including a non vanishing average virtuality $\vpsq \neq 0$ for the fragmenting quark; it will also be necessary to verify that the hard scattering approximator defined in Section~\ref{sec:formalism} allows one to resum the longitudinal gluons into a Wilson line as it happens in ``asymptotic'' factorization theorems \cite{Collins:2007ph,Collins:1989gx,Collins:2011zzd}. The analogies of our scaling variable $\xi_h$ with the $\chi$ variable of the ACOT-$\chi$ heavy quark scheme also deserve further investigation.

Finally, the results presented in this paper point at the necessity of using hadron mass corrected theoretical calculations in QCD fits of fragmentation function that include HERMES and COMPASS data \cite{deFlorian:2017lwf,Borsa:2017vwy,Ethier:2017zbq}, in order to avoid deforming the kaon FFs to compensate for the neglected mass effects. Likewise, other power-suppressed corrections such as Higher-Twist terms in SIDIS should also be included, but this is still, to our knowledge, a largely unexplored topic from a phenomenological point of view.


\begin{acknowledgments}
We thank A.~Bressan, C.-R.~Ji, W.~Melnitchouk, N.~Sato, F.~Steffens, C.~Van Hulse, and C.~Weiss for helpful discussions. This work was supported by the DOE contract No. DE-AC05-06OR23177, under which Jefferson Science Associates, LLC operates Jefferson Lab and DOE contract No. DE-SC0008791.\\
\end{acknowledgments}


\begin{appendix}
  
\section{Treatment of baryon number conservation}
\label{app:baryon_number}

\begin{figure}[htb]
  \centering
  \includegraphics[width=12cm]{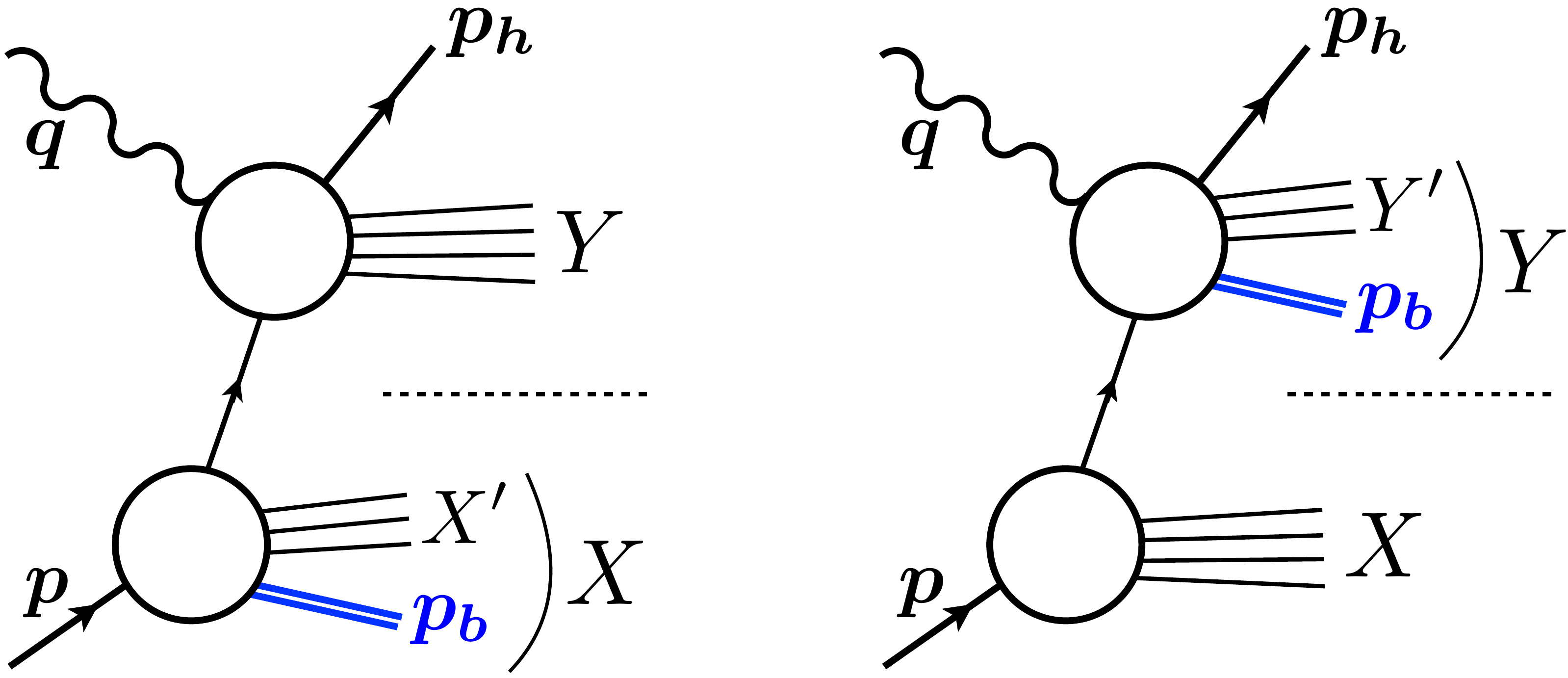}
  \caption{SIDIS in the impulse approximation for the case of an unobserved baryon of momentum $p_b$ produced in the target region (left) and in the current region (right). The separation between these two final state regions, defined as $z_e^{(b)}=0$, is graphically represented by a horizontal dashed line.}
  \label{SIDIS_b_u_d}
\end{figure}

The derivation of the hadron- and nucleon-mass-dependent scaling variables advocated in this work, as well as in Refs.~\cite{Accardi:2009md,Guerrero:2015wha}, relies
on four-momentum and baryon number conservation. In particular, since exactly one baryon is present in the initial state, one baryon, $b$, must also be minimally present in the final state, see Fig.~\ref{SIDIS_b_u_d}.

The scaling variables \eqref{eq:xi_h}-\eqref{eq:zeta_h} have been derived assuming that this baryon is produced predominantly in the target fragmentation region. It is well known that a precise separation of the target and current regions is a subtle matter, as summarized, {\t e.g.}, in Chapter 3 of Ref.~\cite{Niczyporuk:1997jk}. In the present paper and in Ref.~\cite{Guerrero:2015wha}, we take a pragmatic approach and consider a baryon to be produced in the target region if, in analogy with hadron production in electron-positron annihilations, $z_e(b)=-q^2/(2p_b\cdot q) < 0$.
Graphically, this is indicated by the region below the dashed lines in the left panel of Fig.~\ref{SIDIS_b_u_d}. In this case, as was shown in Ref.~\cite{Guerrero:2015wha} and discussed in the main text, four momentum conservation at the hadron level allows one to choose a virtuality $v^2  = 0$ for the incoming quark, but for the scattered quark one needs $\vpsq \geq \frac{m_h^2}{z}$. At LO, where $z=\zeta_h$, this leads to $d\sigma_h  \sim \ q \big(\xi_h \big)  D_q(\zeta_h) $, with  $\xi_h = \Big(1 + \frac{m_h^2}{\zeta_h Q^2} \Big)$.

The assumption just utilized can be heuristically justified by noting that baryon transport in rapidity from the initial to the final state is notoriously difficult, and only about one unit of rapidity is lost by the baryon even in proton-proton scattering \cite{Wong:1995jf}. Therefore, typically, the baryon does not move in rapidity too far away from the target.
Nonetheless, the rapidity gap between the current and target fragmentation region is progressively reduced as $x_B \rightarrow 1$ \cite{Berger:1987zu,Mulders:2000jt}, and the distinction between these becomes blurred. It is thus interesting to explore the kinematics of the case in which the final state baryon appears in the current fragmentation region at $z_e(b)>0$, depicted in the right diagram of Fig.~\ref{SIDIS_b_u_d}.
Following the same arguments as in \cite{Guerrero:2015wha}, one can prove that in this case it is still possible to choose $v^2  = 0$, but that $\vpsq \geq \frac{m_h^2}{z} + \frac{\zeta_h}{z} \frac{M_b^2}{1-\zeta_h}$. At LO, this implies
\begin{equation}
x = \xi_h^{(b)}  \equiv \xi \bigg(1 + \frac{m_h^2}{\zeta_h Q^2} + \frac{M_b^2}{(1-\zeta_h)Q^2} \bigg) \ ,
\label{eq:xi_h_b} \\
\end{equation}
where the superscript denotes that the baryon was produced in the current region, and the parenthesis indicates that it was not observed. Then, the SIDIS cross section for production of a hadron $h$ accompanied by that unobserved baryon,
\begin{equation}
d\sigma_h^{(b)}  \propto  q \big(\xi_h^{(b)} \big)  D_q(\zeta_h) \ ,
\label{eq:CS_b} \\
\end{equation}
is suppressed compared to the case in which the baryon is produced in the target region because $\xi_h^{(b)} > \xi_h$. Numerically, $d\sigma_h^{(b)}$ turns out to be negligible compared to $d\sigma_h$, corroborating the assumption used in the main text.

\vfill

\end{appendix}


\end{document}